\documentclass[twocolumn,aps,amsmath,amssymb]{revtex4}
\usepackage{graphicx}% Include figure files
\usepackage{dcolumn}% Align table columns on decimal point
\usepackage{bm}% bold math
\usepackage {amssymb}
\usepackage {amsmath}

\begin{document}
\title{The effect of magnetic (phase) inhomogeneities upon
the low temperature conductivity of antiferromagnetic
La$_{2}$CuO$_{4+\delta}$ single crystal}

\author{N. V. Dalakova}
\affiliation{B. Verkin Institute for Low Temperature Physics and
Engineering, National Academy of Sciences, Kharkov 61103, Ukraine}

\author{B. I. Belevtsev}
\email{belevtsev@ilt.kharkov.ua} \affiliation{B. Verkin Institute
for Low Temperature Physics and Engineering, National Academy of
Sciences, Kharkov 61103, Ukraine}

\author{E. Yu. Beliayev}
\affiliation{B. Verkin Institute for Low Temperature Physics and
Engineering, National Academy of Sciences, Kharkov 61103, Ukraine}

\author{Yu. A. Kolesnichenko}
\affiliation{B. Verkin Institute for Low Temperature Physics and
Engineering, National Academy of Sciences, Kharkov 61103, Ukraine}

\author{A. S. Panfilov}
\affiliation{B. Verkin Institute for Low Temperature Physics and
Engineering, National Academy of Sciences, Kharkov 61103, Ukraine}

\author{I. S. Braude}
\affiliation{B. Verkin Institute for Low Temperature Physics and
Engineering, National Academy of Sciences, Kharkov 61103, Ukraine}

\begin{abstract}
It is found that a strong inhomogeneous distribution of oxygen in
single crystal La$_{2}$CuO$_{4+\delta}$ can lead to the formation
of isolated superconducting inclusions, the average hole
concentration being no more than 0.0024 per Cu atom. The behavior
of the conductivity in the magnetic field below $T=20$~K is
consistent with possible existence of a insulating low-temperature
magnetic phase (spin density waves), which is typical of cuprates
with excess oxygen.

\end{abstract}

\maketitle

Properties of strongly correlated electronic systems are greatly
affected by local fluctuations and nonuniform distribution of the
charge and spin density. In this respect, the behavior of
high-temperature superconductors (HTSCs) belonging to the family
of lanthanum cuprates can be considered as unique. Such behavior
is often due to the phase separation (typical of these systems)
into regions with different charge-carrier concentrations and the
corresponding local disturbances of structural and magnetic order.
\par
In this study, we revealed a possible effect of spatially
nonuniform distribution of excess oxygen on the low-temperature
behavior of the conductivity of a La$_{2}$CuO$_{4+\delta}$ single
crystal in the antiferromagnetic state. The initial
La$_{2}$CuO$_{4+\delta}$ single crystal had the N\'{e}el
temperature $T_{N}\approx 266$~K and a hole concentration of about
0.0025 per copper atom (according to the estimations based on the
data of Ref. \cite{chen}). X-ray diffraction analysis revealed the
presence of twins, which inevitably arise upon cooling of a
crystal below the point of structural phase transition (530 K)
from the tetragonal phase to the orthorhombic phase \cite{kast}.
At low voltages, the current-voltage ($I$-$V$) characteristics of
this sample have a relatively narrow ohmic portion, in which the
Mott's law of variable-range hopping for three-dimensional systems
is satisfied in the temperature range 20--250 K \cite{mott}:
\begin{equation}
\label{eq1} R \propto \exp\left( {\frac{{T_{0}} }{{T}}}
\right)^{1/4},
\end{equation}

\noindent The localization length $L_c$ can be found from the
$T_0$-value [3]. Our estimate for the CuO$_2$ plane is $L_c\approx
0.4$~nm. At $T < 20$~K, the dependences $R(T)$ deviated from the
Mott's law towards lower resistance.

\par
The initial sample was kept at room temperature in air for 3
years. One might expect the aging of such kind lead to a
significant redistribution of oxygen in the crystal due to the
high mobility of excess oxygen and the tendency of this system to
phase separation. Since holes in La$_{2}$CuO$_{4+\delta}$ are
basically of oxygen origin \cite{kremer}, this process should also
lead to nonuniform distribution of the charge carrier density and
local break of the antiferromagnetic order within and near the
boundaries of hole-enriched regions (arising due to the phase
separation, inherent in doped cuprates \cite{kremer}). The effect
of such inhomogeneities on the transport properties of the
La$_{2}$CuO$_{4+\delta}$ sample is considered below.

\par
As a result of the 3-year exposure of the sample, its N\'{e}el
temperature has increased to 269 K and the hole concentration
lowered to 0.0024. The minor difference in $T_N$ between the
starting and aged samples points to a small decrease in the
concentrations of oxygen and charge carriers. This leads us to
assume that the observed changes in the behavior of the transport
properties are mainly due to the redistribution of oxygen and
charge carriers over the crystal volume rather than to a change in
the average hole concentration.

\begin{figure}[hbt]
\centering\includegraphics[width=0.93\linewidth]{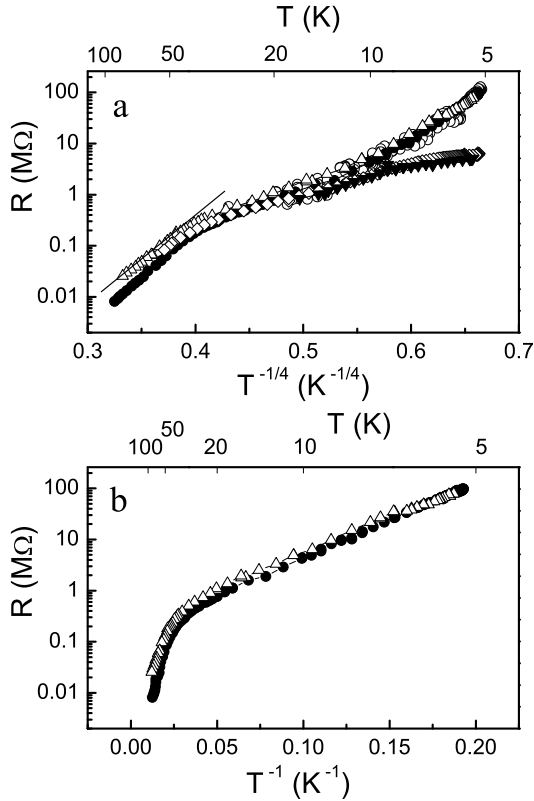}
\caption{(a) Temperature dependence of the resistance $\lg R$ {\it
vs.} $T^{-1/4}$ of a La$_{2}$CuO$_{4+\delta}$ single crystal
($T_{N}\approx 269$~K) at different measuring currents: ($\circ$)
0.03, ($\bullet$) 0.06, ($\vartriangle)$ 0.2, ($\diamondsuit$)
0.5, and ($\blacktriangledown$) 2$\mu$A; (b) the dependence $\lg
R$ {\it vs.} $T^{-1}$ for currents of ($\bullet$) 0.06 and
($\vartriangle$) 0.2 $\mu$A. }
\end{figure}
\begin{figure}[t]
\centering\includegraphics[width=0.93\linewidth]{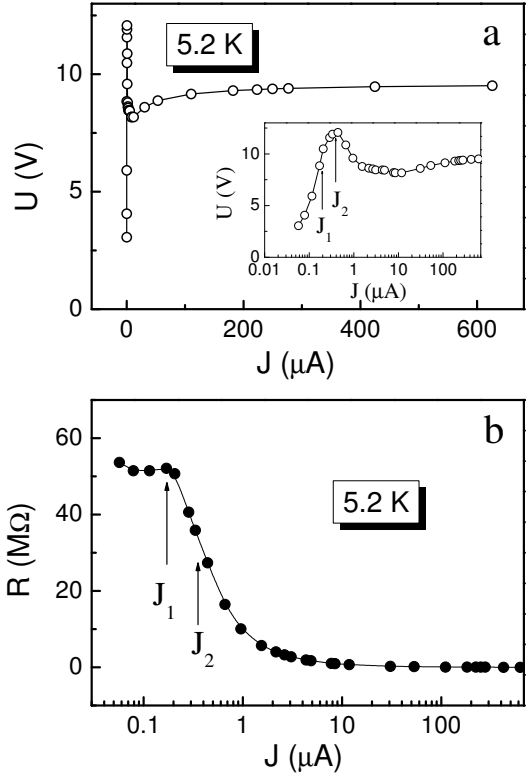}
\caption{(a) Current-voltage characteristic of the
La$_{2}$CuO$_{4+\delta}$ single crystal at $T=5.2$~K. Inset: the
same dependence in the semilogarithmic coordinates; (b) the
corresponding dependence of the resistance on current.}
\end{figure}

\par
The resistance along CuO$_2$ planes was measured with a constant
current through the sample. The current was varied from 0.03 to 2
$\mu$A. The temperature dependences $R(T)$ in the range 5-100 K at
several currents are shown in Figs. 1a and 1b. As the temperature
lowers to ~ 45 K, the curves deviate from the Mott's law (Fig.1a)
towards lower resistance. At $T < 25$~K and $J\leq 0.2$ $\mu$A
there is a change to a simple activation dependence $R\propto
\exp(\Delta/kT)$, where $\Delta =32.4$~K (Fig.1b). Above 45~K, the
dependence $R(T)$ at a current 0.2~$\mu$A is described by Eq. (1)
with $T_{0}\approx 2.3\times 10^{6}$~K, which corresponds to the
localization length $L_c\approx 0.262$~nm. The obtained value of
$L_c$ is much smaller than the orthorhombic lattice parameters in
a CuO$_2$ plane ($a = b = 0.54$~nm). This points out to a much
stronger localization and less uniform distribution of charge
carriers in comparison with the initial state.
\par
The behavior of $R_{ab}(T)$ in the temperature range $T<50$~K is
nonohmic; it demonstrates a strong dependence of the resistance on
current. At some critical current $J>0.2$~$\mu$A, the resistance
sharply decreases (Fig. 1a). The changes in $R(T,J)$ correlate
with the behavior of the $I$-$V$ characteristics in the
temperature range 5-80 K. Typical behavior of the $I$-$V$
characteristics and the corresponding dependences of the
resistance on current is shown in Figs. 2a and 2b. The dependences
$U(J)$ and $R(J)$ are essentially nonlinear and have a region with
current-controlled negative differential resistance. The
characteristic currents $J_1$ and $J_2$ are marked with arrows.
The current $J_1$ corresponds to the inflection point in the
ascending portion of the dependence $U(J)$ (Fig. 2a, inset) and
the crossover point in the curve $R(J)$ (Fig. 2b). The current
$J_2$ corresponds to the maximum in the dependence $U(J)$ and the
inflection point in the lower branch of the curve $R(J)$ . These
features of the $I$-$V$ characteristics are observed at all
temperatures in the range 5-80 K. Comparison of the $I$-$V$
characteristics with the temperature dependences of the resistance
shows that the character of the dependence $R(T)$ changes at $J =
J_1$. A simple exponential dependence is observed at currents
$J\leq J_1$ ($J_1 = 0.2$~$\mu$A). Deviation from the law $R\propto
\exp(1/T)$ (see Fig. 2) occurs at currents $J > J_2$ ($J_2 =
0.44$~$\mu$A).

\begin{figure}[t]
\centering\includegraphics[width=0.93\linewidth]{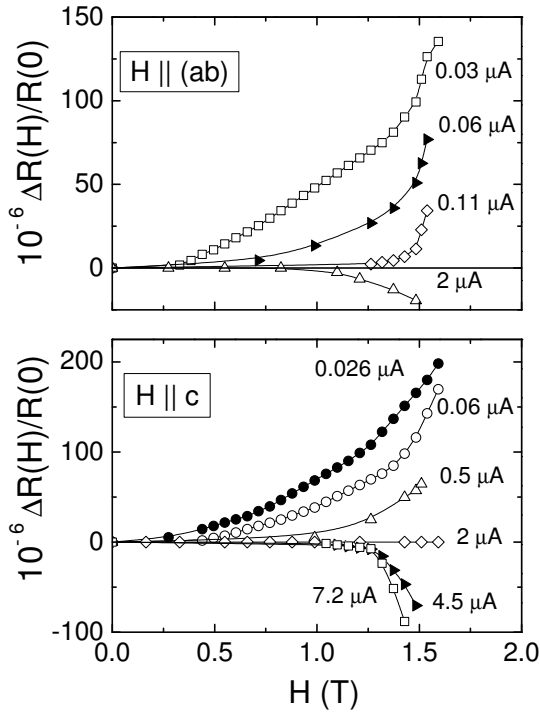}
\caption{Magnetoresistance  curves taken at $T=5$~K for different
transport currents $J\parallel (ab)$ for two the magnetic field
directions: $\textbf{H}\parallel (ab)$ and $\textbf{H}\parallel
\textbf{c}$.}
\end{figure}

\par
The behavior of the magnetoresistance in the
La$_{2}$CuO$_{4+\delta}$ single crystal studied is also dependent
on the transport current in the range 5-80 K. The
magnetoresistance is positive for low currents in the range $T\leq
17$~K. Positive magnetoresistance is observed at temperatures and
currents for which the law $R\propto \exp(\Delta/kT)$ is obeyed.
The magnetoresistance becomes negative with an increase in the
temperature or current (Fig. 3). Above 17 K, only negative
magnetoresistance is observed, whose magnitude decreases with an
increase in current.
\par
It is known that, in the regime of hopping conductivity in
sufficiently low electric fields ($E\ll kT/er_{h}\gamma$), the
resistance is field-independent (here, $r_{h}$ is the mean hopping
distance and $\gamma$ is a numerical factor of the order of
unity). According to the estimation performed, this is true to a
great extent for the sample under study; therefore, the observed
nonlinear behavior of the $I$-$V$ characteristic (Fig. 2) cannot
be directly attributed the effect of electric field on the hopping
conductivity of a homogeneous system. At the same time, the weak
dependence of the resistance on current at $J\geq 1$~$\mu$A
suggests the absence of Joule heating.
\par
The dependence  $R\propto \exp(\Delta/kT)$ observed in the low
temperature range suggests that the oxygen disordering produces a
gap ($\Delta =32.4$~K) in the spectrum of quasiparticle
excitations involved in the charge transfer. It is interesting to
clarify the nature of this gap.
\par
The revealed changes in the behavior of hopping conductivity below
25 K correspond well to the known phenomena in inhomogeneous
systems (mixtures of insulating and superconducting phases).
Cooling of such a system below $T_c$ causes a transition from the
variable-range hopping to the dependence $R\propto
\exp(\Delta/kT)$, where $\Delta$ is a superconducting gap, playing
the same role as the band gap in insulators \cite{boris1,efet}. In
our case, the formation of a superconducting phase can be related
to the phase separation in lanthanum cuprates \cite{kremer,well},
which yields two phases with $\delta  =0$ and $\delta >0$.
Previously, the transition to a simple exponential dependence in
La$_{2}$CuO$_{4+\delta}$, caused by nonuniform oxygen
distribution, was observed at $T < 20$~K for a far less resistive
sample of La$_{2}$CuO$_{4+\delta}$ with $L_c\approx 1$~nm
\cite{boris2} and a hole concentration of 0.0044. The sample
studied has a much lower average hole concentration and a shorter
localization length (0.26 nm). At the same time, it is
characterized by a larger degree of inhomogeneity due to the more
nonuniform oxygen distribution as a result of the long-term
exposure (aging). In our opinion, it is the phase separation that
determines the observed features of low-temperature behavior of
the resistance of the sample studied. With a decrease in
temperature, the oxygen-enriched isolated regions turn to the
superconducting state. As a result of the decrease in the density
of single-particle excitations and weak coupling between isolated
superconducting inclusions at sufficiently low temperatures, a
transition from dependence (1) to a stronger simple exponential
dependence occurs. An increase in current leads to depairing of
carriers and, possibly, their heating under conditions of strong
inhomogeneity. Either of the mechanisms should lead to a decrease
in the resistance, which is observed at currents $J > 0.2$~$\mu$A.
\par
The suppression of local superconductivity by the magnetic field is 
also expected to break the pairs of charge carriers and reduce 
resistance \cite{boris1}. We however observed positive magnetoresistance 
in the low temperature region at quite small currents, which is rather 
unexpected result. This behavior of magnetoresistance can be attributed 
to the new low-temperature magnetic phase recently detected in 
La$_{2}$CuO$_{4+\delta}$. The investigation
of phase separation in lanthanum cuprates with excess oxygen shows that the 
dielectric phase in the mixed system below $T=40$~K is a new magnetic 
state, namely, a spin-density wave \cite{manot} rather than the N\'{e}el phase. 
The relative volume of this phase increases with a decreasing in content of 
excess oxygen (with correspondin decrease in $T_c$). Therefore, in 
lightly doped La$_{2}$CuO$_{4+\delta}$ compound, this phase is rather abundant 
and the content of the superconducting phase is much smaller. The magnetic field 
stabilizes the magnetic phase by reducing the volume fraction of 
the superconducting phase and thus imitates a change in the doping level 
\cite{manot}. The effect is not yet clear completely but it is 
obvious that in this case the magnetic field should induce positive 
magnetoresistance.
\par
In conclusion, the results of this investigation indicate that, under conditions 
of strong inhomogeneity of a sample at temperatures $T < 25$~K, phase separation 
even in La$_{2}$CuO$_{4+\delta}$ with a low content of excess oxygen (hole
concentration not higher than 0.0024 per copper atom) may lead to the formation 
of isolated regions (clusters) of a superconducting phase. 
The magnetoresistance behavior of the sample in this temperature interval 
can be attributed to the presence of the insulating magnetic phase (spin density wave), 
as assumed in \cite{manot}.

\end{document}